\documentclass[twocolumn,apsfloatfix]{revtex4}
\usepackage[pdftex]{graphicx,color}
\usepackage{amsmath,amssymb,wasysym,graphicx,capt-of,ifthen,calc}
\usepackage{enumerate}
\usepackage{subfigure}
\usepackage{float}
\begin{document}

\title{Dynamics of Voter Models on Simple and Complex Networks} 
\author{S.~Redner}
\affiliation{Santa Fe Institute, 1399 Hyde Park Road, Santa Fe, NM, 87501
  USA}

\maketitle

\section{Introduction}\label{sec:intro} 

How do groups of people come to consensus?  While it's hard to imagine a
large group being able to agree on anything, there are some settings where
unanimity is necessary---juries are one example.  The voter model (VM)
represents an idealization of this opinion evolution in which each
individual, or agent, is influenced only by other members of the group; there
is also no notion of a ``right'' or a ``wrong'' opinion, and there are no external
influences, such as news media.  In the VM, each agent, or voter, can assume
one of two states (e.g., $\mathbf{0}$/$\mathbf{1}$, normal/mutant,
Democrat/Republican).  One agent resides at each node of a lattice or an
arbitrary network and updates its state at unit rate until a population of
$N$ agents necessarily reaches consensus.

\begin{figure}[ht]
\centerline{  \includegraphics[width=0.35\textwidth]{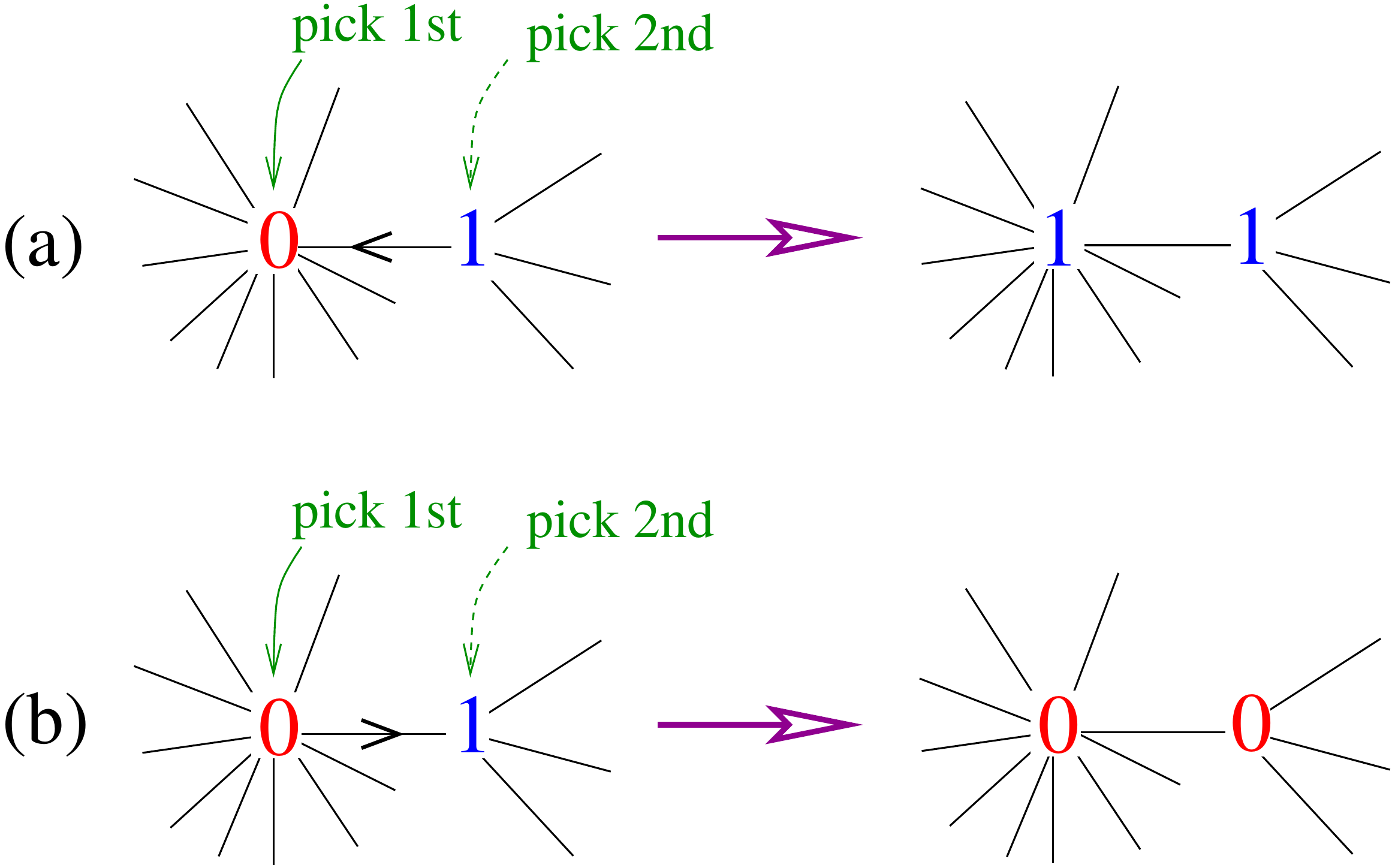}}
\caption{Update rules.  (a) VM: a randomly chosen agent adopts the state of a
  random neighbor; (b) IP: a randomly chosen agent exports its state to a
  neighbor. }
\label{model}
\end{figure}

In detail, VM evolution is as follows (Fig.~\ref{model}):
\begin{itemize}
\itemsep -1ex
\item Pick a random node (a voter).
\item The voter adopts the state of a random neighbor.
\end{itemize}
An anthropomorphic interpretation of VM dynamics is that each agent has zero
self-confidence and merely adopts one of its neighbor's state.  We also
discuss the related invasion process (IP), whose update rule is:
\begin{itemize}
\itemsep -1ex
\item Pick a random node (an invader).
\item The invader exports its state to a random neighbor.
\end{itemize}
Pictorially, an invader dictatorially imposes its state on one of its
neighbors; equivalently, an invader replicates and its offspring replaces an
neighboring agent.  While the differences between these two models appear
superficially trivial, they are fundamental on complex networks.  The two
basic questions that we will discuss are: (i) What is the time to reach
consensus?  (ii) By what route is consensus achieved?  We present our
collaborative work on this subject with Vishal Sood and Tibor
Antal~\cite{SAR}.

\section{Conservation Laws}\label{sec:model}

A crucial feature of VM and IP dynamics on arbitrary networks is that they
satisfy conservation laws that determine their long-time behaviors.  Let's
develop the language to uncover these laws~\cite{L,K92}.  Define $\eta$ as
the state of the entire network, and $\eta(x)$, which can equal $\mathbf{0}$
or $\mathbf{1}$, as the state of node $x$.  In an update, the node state
changes from $\mathbf{0}$ to $\mathbf{1}$ or vice versa.  Let
$\mathbf{\eta}_{x}$ denote the network state after the node at $x$ changes
state.  We may succinctly write the transition probability that node $x$
changes state as
\begin{equation}
  \label{master}
  \textbf{P} [\eta \to \eta_{x}] = 
  \sum_{y} \frac{A_{x y}}{N\mathcal{Q}}\,[\Phi(x,y)+\Phi(y,x)],
\end{equation}
where $\Phi(x,y)\equiv \eta(x)[1-\eta(y)]=1$ if the states at $x$ and $y$
differ and $\Phi(x,y)= 0$ if these states agree, and $A_{x y}$ is the
adjacency matrix.  Although \eqref{master} looks formidable, its meaning is
simple: $A_{x y}\,[\Phi(x,y)+\Phi(y,x)]$ is non-zero only when nodes $x$ and
$y$ are connected and in opposite states, so that an update actually occurs.
For the VM, the factor $(N\mathcal{Q})^{-1}\!=\! (Nk_x)^{-1}$ accounts for
first choosing any node $x$ with probability $1/N$, and then one of its
neighbors $y$ with probability $1/k_x$, where $k_x$ is the degree of node
$x$.  In the IP, $(N\mathcal{Q})^{-1}\!=\! (Nk_y)^{-1}$: first choose node
$y$ (a neighbor of $x$) with probability $1/N$, and then choose $x$ with
probability $1/k_y$.

The kernel for the evolution of the population is the average change in the
state of a single node, $\langle \Delta\eta(x)\rangle$.  This change equals
the probability that $\eta(x)$ changes from $\mathbf{0}$ to $\mathbf{1}$
minus the probability of a change from $\mathbf{1}$ to $\mathbf{0}$:
$\langle\Delta\eta(x)\rangle = \left[1-2\eta(x)\right]
\mathbf{P}[\eta\rightarrow\eta_x]$.
Summing this transition probability over all nodes gives the average change
in $\rho$, the density of nodes in state $\mathbf{1}$:
\begin{equation}
\label{drho-av}
\langle \Delta \rho\rangle = \sum_x \langle \Delta \eta(x)\rangle =   
\sum_{x,y} \frac{A_{x y}}{N\mathcal{Q}} \left[\eta(y)-\eta(x)\right]\,.
\end{equation}

Since $\mathcal{Q}$ is constant on regular lattices, the summand on the right
is antisymmetric in $x$ and $y$ and $\langle \Delta\rho\rangle=0$.  Thus
$\langle\rho\rangle$ \emph{is conserved}.  This innocuous-looking
conservation law has far-reaching consequences.  It immediately gives the
\emph{fixation} or \emph{exit probability} namely, the probability
$\mathcal{E}(\rho)$ that a finite system with an initial density $\rho$ of
$\mathbf{1}$s attains consensus of $\mathbf{1}$s.  Because $\rho$ is
conserved and because the final state consists of either all $\mathbf{1}$s or
all $\mathbf{0}$s, we have
$\rho=\mathcal{E(\rho)}\cdot \mathbf{1}+
[1-\mathcal{E}(\rho)]\cdot\mathbf{0}$.
Thus with no calculation the fixation probability equals $\rho$\,!

The power of this conservation law suggests looking for analogous laws for
the VM and the IP on degree-heterogeneous networks.  To obtain a conserved
quantity, the factor $\mathcal{Q}$ in the denominator of the transition rate
in \eqref{master} must somehow be canceled out.  This leads us to generalize
the notion of density to the {\em degree-weighted moments}
$ \omega_{m} \equiv \sum_k k^m n_k \rho_k/\mu_{m}$ (note that $\omega_0=\rho$
and for simplicity we write $\omega_1$ as $\omega$), where
$\rho_k \equiv \sideset{}{'}\sum_{\!\!x}\! \eta(x)/N_k$ is the density of
$\mathbf{1}$s on the subset of nodes of degree $k$, the prime restricts the
sum to nodes $x$ of degree $k$.  Here $\mu_m= \sum_k k^m n_k$ is the
$m^{\rm th}$ moment of the degree distribution of the network, with $N_k$
($n_k$) the number (density) of nodes of degree $k$.  Repeating the
calculation in Eq.~\eqref{drho-av} for $\langle\omega\rangle$ for the VM and
for $\langle\omega_{-1}\rangle$ for the IP, it is immediate to show that the
conserved quantities are:
\begin{align}
\begin{split}
\label{cons}
&\langle\omega\rangle   \qquad \qquad ~~\,\, \text{VM},\\
& \langle\omega_{-1}\rangle   \qquad\qquad\text{IP}.
\end{split}
\end{align}
Since the initial value of the conserved quantity equals its value in the
final unanimous state, the exit probability is
\begin{align}
\begin{split}
\label{exit-cons}
&\mathcal{E}(\omega)=\omega ~~~~~\,\, \qquad \qquad \text{VM},\\
&\mathcal{E}(\omega_{-1})=\omega_{-1}  \qquad \qquad \text{IP}.
\end{split}
\end{align}
An instructive example is the star graph, where $N$ nodes are connected only
to a single central hub.  For the VM, if the hub is in state $\mathbf{1}$ and
all other nodes are in state $\mathbf{0}$, then \eqref{exit-cons} mandates
that the probability of reaching $\mathbf{1}$ consensus is 1/2\,!  That is, a
single well-connected agent largely determines the final state.  Conversely,
in the IP, a mutant at the hub is very likely to be extinguished (fixation
probability $\propto N^{-2}$), while a mutant at the periphery is more likely
to persist (fixation probability $\propto N^{-1}$).

\section{VOTER MODEL ON NETWORKS}
\label{sec:vm-cg}

\subsection{Complete Graph}

To understand the VM and the IP on complex networks, first consider the
complete graph, where the VM and the IP are identical.  In each update event,
$\rho\to\rho\pm \delta\rho$, with $\delta\rho=1/N$, corresponding to a voter
undergoing the respective state changes $\mathbf{0}\to\mathbf{1}$ or
$\mathbf{1}\to\mathbf{0}$.  The probabilities for these respective events are:
\begin{align}
\label{rl}
\begin{split}
  \mathbf{R}(\rho)&\equiv \mathbf{P}[\rho\rightarrow\rho+\delta \rho] = 
(1-\rho)\rho \\
  \mathbf{L}(\rho)&\equiv \mathbf{P}[\rho\rightarrow\rho -\delta \rho] = 
\rho(1-\rho)\,.
\end{split}
\end{align}
We term $\mathbf{R}$ and $\mathbf{L}$ as the raising and lowering operators.

We now use these transition probabilities to write the evolution equation for
the average time $T(\rho)$ to reach consensus when the fraction of agents
initially in state $\mathbf{1}$ is $\rho$ (the \emph{backward Kolmogorov
  equation}~\cite{VK,R01}):
\label{TTT}
\begin{eqnarray}
\label{T}
  T (\rho)&= \delta t +  \mathbf{R}(\rho)T(\rho\!+\!\delta\rho)
 +\mathbf{L}(\rho) T(\rho\!-\!\delta\rho) \nonumber\\
&+ [1-\mathbf{R}(\rho)-\mathbf{L}(\rho)] T(\rho)\,.
\end{eqnarray}
This simple-looking, but deceptively powerful equation expresses the average
consensus time as the time $\delta t$ for a single update step plus the
average time to reach consensus after this update.  The three terms account
for the transitions $\rho \rightarrow \rho\pm\delta\rho$ or
$\rho \rightarrow \rho$, respectively.  Expanding Eq.~\eqref{T} to second
order in $\delta \rho$ gives
\begin{equation}
\label{T-back}
 v(\rho)\frac{d T(\rho)}{d\rho}+D(\rho)\frac{d^2 T(\rho)}{d \rho^2} = -1\,,
\end{equation}
with drift velocity $v(\rho)\propto [\mathbf{R}(\rho)-\mathbf{L}(\rho)]$ and
diffusivity $D(\rho)\propto [\mathbf{R}(\rho)+ \mathbf{L}(\rho)]$.  On the
complete graph, the drift term is zero and only the diffusion term, which
quantifies the stochastic noise, remains.  For the boundary conditions
$T(0)\!=\! T(1)\!=\! 0$ (consensus time equals 0 if the initial state is
consensus) the solution is
\begin{equation}
\label{VM:KG:contime:sol}
  T(\rho)= -N\big[(1-\rho)\ln(1-\rho)+\rho\ln \rho\big] \,.
\end{equation}
For equal initial densities of each opinion, $T\big(\frac{1}{2}\big)=N\ln 2$,
while for a single mutant, $T\big(\frac{1}{N}\big)\approx \ln N$.  The linear
dependence on $N$ represents the generic behavior for the consensus time of
the VM on Euclidean lattices in spatial dimensions $d\geq 3$.

\subsection{Complete Bipartite Graph\label{VM:Sec:BG}}

\begin{figure}[ht]
\centerline{\includegraphics[width=0.3\textwidth]{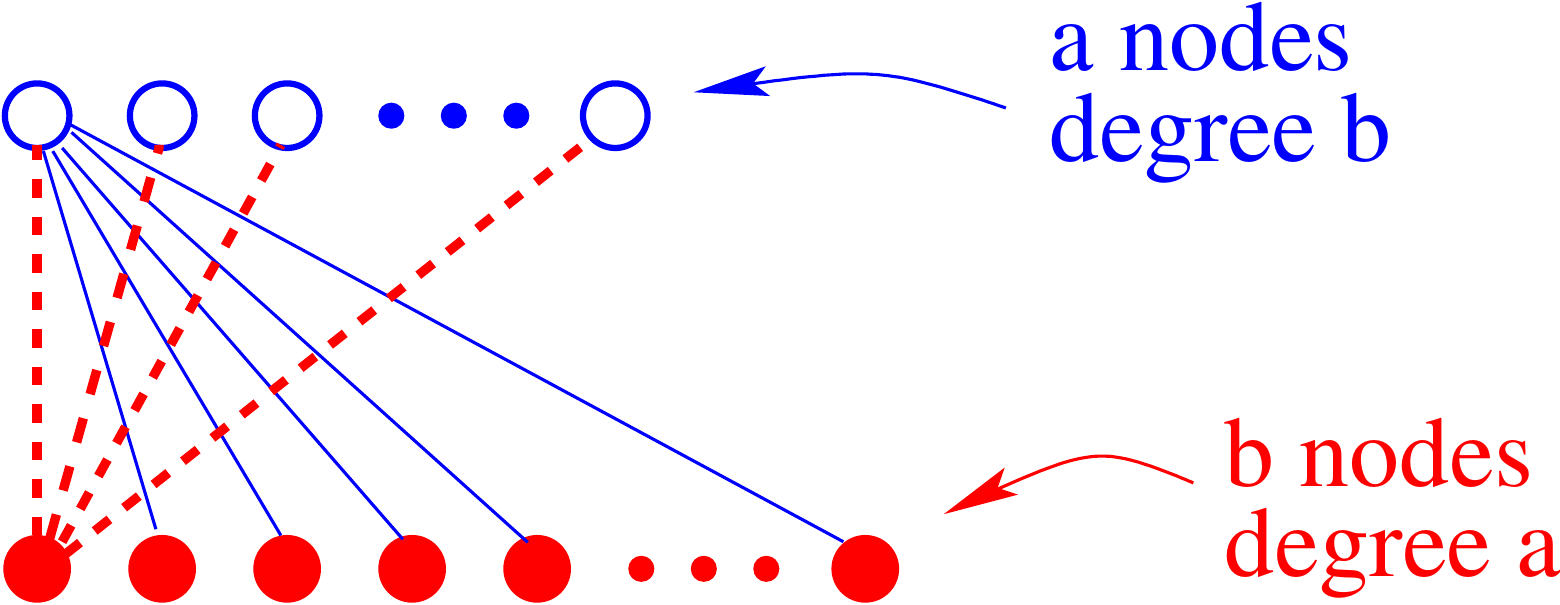}}
\caption{The complete bipartite graph $K_{a, b}$.}
\label{Kab}
\end{figure}

An important clue to understanding how degree heterogeneity affects the
dynamics is provided by studying the simplest network network that contains
of nodes with different degrees---the complete bipartite graph $K_{a,b}$.  In
this graph, $a+b$ nodes are partitioned into two subgraphs of size $a$ and
$b$ (Fig.~\ref{Kab}).  Each node in subgraph $\mathbf{a}$ links to all nodes
in $\mathbf{b}$, and vice versa.  Thus $\mathbf{a}$ nodes all have degree
$b$, while $\mathbf{b}$ nodes all have degree $a$.

We can immediately determine the exit probability by using the conservation
law from Eq.~\eqref{cons},
$\langle\omega\rangle=\frac{1}{2} (\rho_\mathbf{a}+\rho_\mathbf{b})$.  For
example, when one subgraph contains only $\mathbf{0}$s and the other only
$\mathbf{1}$s, the probability to reach $\mathbf{1}$ consensus is
$\frac{1}{2}$, \emph{independent} of the $\mathbf{a}$ and $\mathbf{b}$
subgraph sizes.

To determine the dynamical behavior, let $N_{\mathbf{a},\mathbf{b}}$ be the
respective number of voters in state $\mathbf{1}$ on each subgraph, with
$\rho_\mathbf{a} = N_\mathbf{a}/a$, $\rho_\mathbf{b} = N_\mathbf{b}/b$ the
respective subgraph densities.  In an update, these densities change
according to the raising/lowering transition probabilities,
\begin{equation*}
  \label{VM:BG:jump} 
\begin{split}
  \mathbf{R}_\mathbf{a} \equiv \mathbf{P}[\rho_\mathbf{a},\rho_\mathbf{b} 
\rightarrow \rho_\mathbf{a}^+,\rho_\mathbf{b}] 
= \frac{a}{a+b}\,\,\rho_\mathbf{b} (1-\rho_\mathbf{a}), \\
  \mathbf{L}_\mathbf{a} \equiv \mathbf{P}[\rho_\mathbf{a},\rho_\mathbf{b}
 \rightarrow \rho_\mathbf{a}^-, \rho_\mathbf{b}] 
= \frac{a}{a+b}\,\, \rho_\mathbf{a} (1-\rho_\mathbf{b}), 
\end{split}
\end{equation*}
with $\rho_\mathbf{a}^\pm=\rho_\mathbf{a}\pm a^{-1}$.  Here
$\mathbf{R}_\mathbf{a}$ is the probability to increase the number of
$\mathbf{1}$s in subgraph $\mathbf{a}$ by 1, for which we need to first
choose an agent in state $\mathbf{0}$ in $\mathbf{a}$ and then an agent in
state $\mathbf{1}$ in $\mathbf{b}$.  Similarly, $\mathbf{L}_\mathbf{a}$ gives
the corresponding the probability for reducing the number of $\mathbf{1}$s in
$\mathbf{a}$.  Analogous definitions hold for $\mathbf{R}_\mathbf{b}$ and
$\mathbf{L}_\mathbf{b}$ by interchanging $a\leftrightarrow b$.

\begin{figure}[ht]
\centerline{
\subfigure[]{\includegraphics[width=0.22\textwidth]{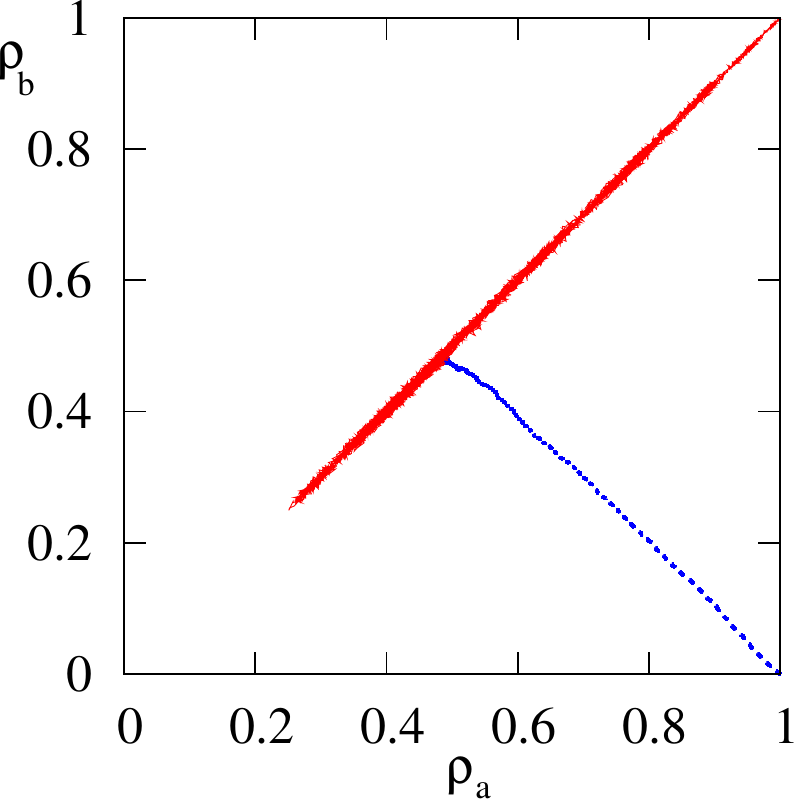}}
\subfigure[]{\raisebox{1mm}{\includegraphics[width=0.215\textwidth]{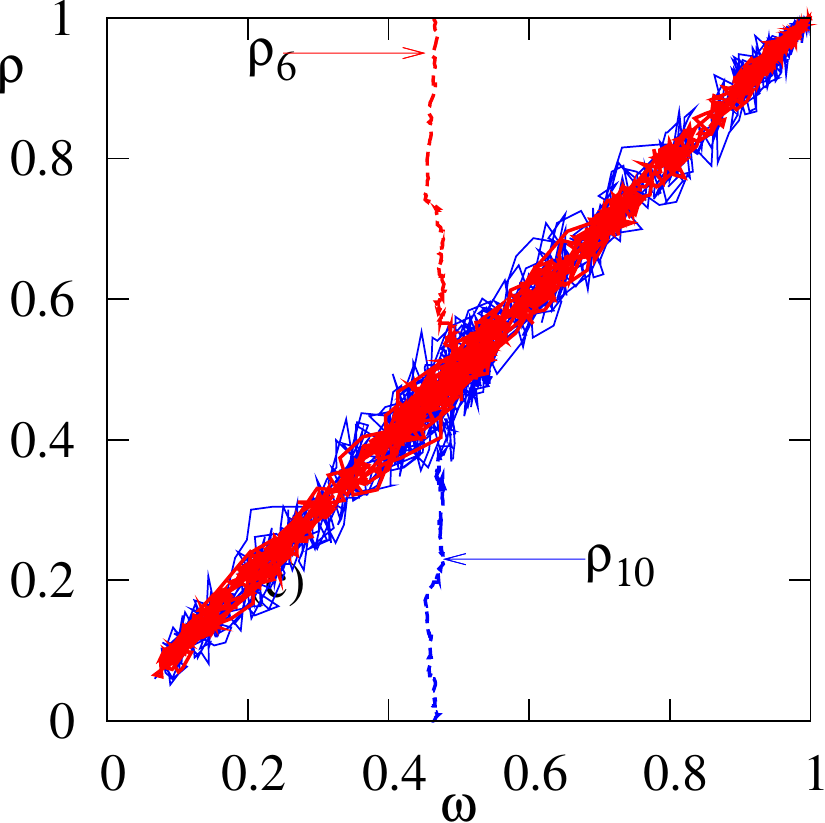}}}
}
\caption{Evolution of subgraph densities for single VM realizations on: (a) a
  complete bipartite graph of $2\times 10^5$ nodes (with $a=b=10^5$), and (b)
  a configuration model of $2\times 10^5$ nodes with degree distribution
  $n_k\sim k^{-2.5}$.  In (a), the dotted curve is the transient from the
  initial state $(\rho_\mathbf{a},\rho_\mathbf{b})=(1,0)$ before the slow
  approach to consensus after $\approx 10^4$ time steps (solid).  In (b), the
  initial state is $(\rho_{k>\mu_1},\rho_{k\leq\mu_1}) =(0,1)$.  Shown are
  $\rho_{6}(t)$ (degree less than $\mu_1=8$) and $\rho_{10}(t)$ (degree
  greater than $\mu_1$) versus $\omega$.  The initial transient lasts
  $\approx 2$ time steps (dotted), while consensus occurs after 1742 time
  steps. }
\label{VM:BG:Fig:Evo}
\end{figure}

From these transition probabilities, the rate equations for the average
subgraph densities are
$\dot\rho_{\mathbf{a}}= \rho_{\mathbf{b}}-\rho_{\mathbf{a}}$ and
$\dot\rho_{\mathbf{b}}= \rho_{\mathbf{a}}-\rho_{\mathbf{b}}$.  Their
solutions show that the subgraph densities are driven to the common value
$\frac{1}{2}[\rho_\mathbf{a}(0)+ \rho_\mathbf{b}(0)]$ in a time of order 1
(Fig.~\ref{VM:BG:Fig:Evo}(a)).  Thus the total density of $\mathbf{1}$s,
which evolves as
$\dot \rho = \left(a\dot\rho_\mathbf{a} + b\dot\rho_\mathbf{b}\right)(a+b)$,
becomes conserved in the long-time limit.  Therefore, there is a two
time-scale approach to consensus: initially, the effective bias quickly
drives the system to equal subgraph densities
$\rho_\mathbf{a}=\rho_\mathbf{b}$; subsequently, diffusive fluctuations drive
the population to consensus.  This dynamical picture also arises for general
complex networks.

To determine the consensus time $T(\rho_\mathbf{a},\rho_\mathbf{b})$, we
exploit the feature that $\rho_\mathbf{a}\to \rho_\mathbf{b}$ in the
long-time limit.  Then following exactly the same steps as those for the
complete graph, the consensus time satisfies
\begin{equation}
\label{VM:BG:CT:omega:eq}
  \omega (1-\omega)\frac{\partial^2T}{\partial\omega^2} =-\frac{4ab}{a+b}\,,
\end{equation}
with solution, for $T(0)=T(1)=0$,
\begin{equation}
\label{VM:BG:CT:sol}
T(\omega)= \frac{4 a b}{a+b} \left[(1-\omega)\ln(1-\omega)
  +\omega\ln\omega\right] .
\end{equation}
The consensus time has the same form as in the complete graph
[Eq.~(\ref{VM:KG:contime:sol})], but with an effective population
$N_{\mathrm{eff}} = 4ab/(a+b)$.  If both the $\mathbf{a}$ and $\mathbf{b}$
subgraphs have similar sizes, $a, b\approx N/2$, then $N_{\rm eff}\approx N$.
However, if, for example, $a\sim \mathcal{O}(1)$ and $b\approx N$ then
$T\sim \mathcal{O}(1)$\,!  One highly-connected node can promote consensus.

\subsection{Complex Networks}
\label{subsec:HDN}

Now we turn to VM and IP dynamics on complex networks.  While the bookkeeping
becomes a bit tedious, the approach is morally the same as that for the
complete bipartite graph: separate the dynamics according to the degree of
each node.  From Eq.~\eqref{master}, the transition probabilities for
increasing and decreasing the density of voters of type $\mathbf{1}$ on nodes
of fixed degree $k$ are:
\begin{equation}
  \label{VM:HG:jump} 
\begin{split}
  \mathbf{R}_k[\{\rho_k\}] &\!\!\equiv\!\! \mathbf{P}[\rho_k\rightarrow\rho_k^+]
  = \frac{1}{N} \sideset{}{'}\sum_{x,y}\!\frac{1}{k_{x}}\, A_{xy}\,\Phi(y,x)\\
  \mathbf{L}_k [\{\rho_k\}] &\!\!\equiv\!\! \mathbf{P}[\rho_k\rightarrow\rho_k^-]
  = \frac{1}{N} \sideset{}{'}\sum_{x,y}\!\frac{1}{k_{x}}\, A_{xy}\, \Phi(x,y),
\end{split}
\end{equation}
where $\rho_k^{\pm}=\rho_k\pm N_k^{-1}$, and the prime restricts the sum to
nodes of fixed degree $k$.  In this equation, the densities associated with
nodes of degrees $k'\ne k$ are unaltered.

We now make the simplification of considering the mean-field
\emph{configuration model} (see, e.g.,~\cite{N}).  This is a network that is
constructed by starting with a set of nodes that have ``stubs'' of specified
degrees, and then connecting the ends of stubs at random until no free ends
remain.  By this construction, the degrees of neighboring nodes are
uncorrelated.  Thus we may replace $A_{xy}$ by
$\langle A_{xy}\rangle = k_xk_y/\mu_1N$ in \eqref{VM:HG:jump}.  Following the
same steps as in the complete bipartite network, the backward Kolmogorov
equation for the consensus time is
\begin{equation}
\label{VM:Kol-t}
  \sum_k  v_k \frac{\partial T}{\partial\rho_k} +
  \sum_k  D_k \frac{\partial^2 T}{\partial\rho_k^2} =-1, 
\end{equation}
with degree-dependent velocity and diffusivity ($v_k,D_k)$.

\begin{figure}[ht]
\centerline{\includegraphics[width=0.325\textwidth]{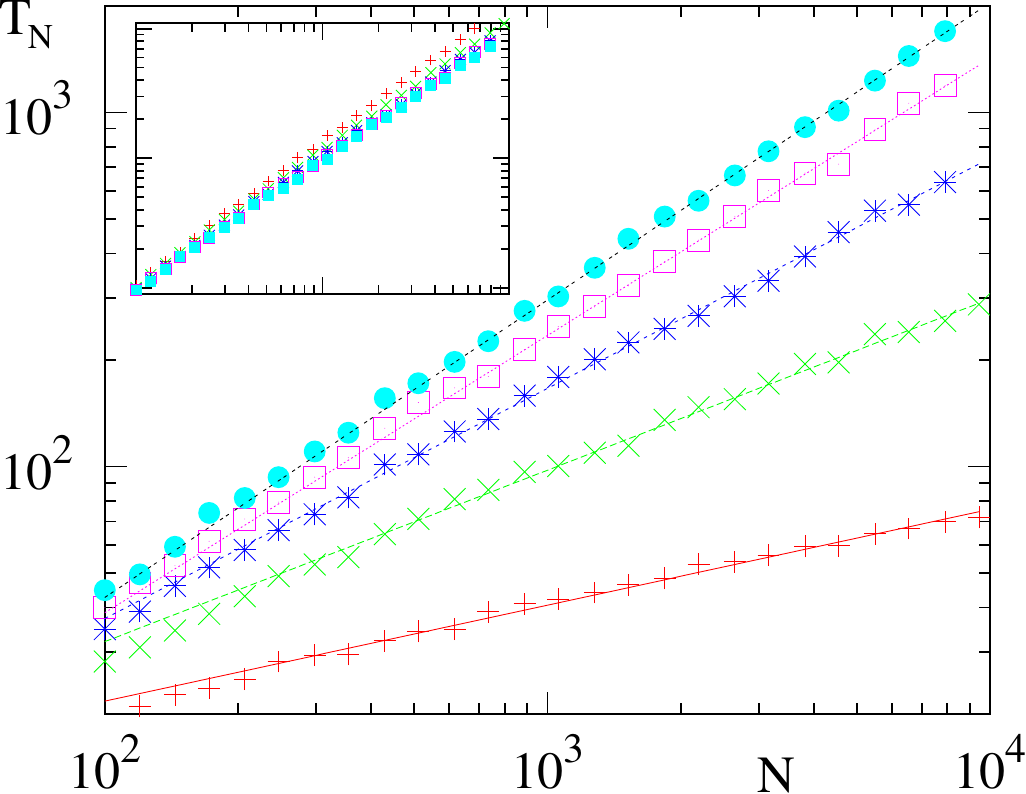}}
\caption{Consensus time $T_N$ versus $N$ for the configuration model with
  degree distribution $n_k = k^{-\nu}$ for $\nu = 2.1$ $(+)$, $2.3$
  $(\times)$, $2.5$ $(\ast)$, $2.7$ $(\circ)$ and $2.9$ $(\bullet)$.  Data
  are based on 100 graph realizations and 10 realizations of VM dynamics on
  each graph.  Lines represent the prediction \eqref{VM:HG:SF:CT}.  The inset
  shows the same data plotted in the scaled form $\mu_2T_N/\mu_1^2$ versus
  $N$.}
\label{TCU}
\end{figure}

To simplify \eqref{VM:Kol-t}, it is helpful to first study the time
dependence of the density of voters in state $\mathbf{1}$ on nodes of fixed
degree $k$.  As seen in Fig.~\ref{VM:BG:Fig:Evo}(b) (and can be shown
analytically) the average densities $\langle \rho_k\rangle$ all converge to
the common value $\omega$ in a time of the order of 1.  Thus at long times,
$v_k$ in \eqref{VM:Kol-t} vanishes.  We also convert derivatives with respect
to $\rho_k$ to derivatives with respect to $\omega$ by
$\frac{\partial T}{\partial\rho_k}=\frac{\partial T}{\partial\omega}
\frac{\partial\omega}{\partial\rho_k} = \frac{k n_k}{\mu_1}\frac{\partial T}{\partial\omega}\,,$
to reduce \eqref{VM:Kol-t} to
\begin{equation}
\label{VM:HG:BPE}
 \frac{\mu_2}{N\mu_1^2}\, \omega(1-\omega)\, 
\frac{\partial^2 T}{\partial \omega^2} =-1. 
\end{equation}
Defining an effective population size by
$N_{\mathrm{eff}} = N\,{\mu_1^2}/{\mu_2}$, and comparing with
\eqref{VM:BG:CT:omega:eq}, the consensus time is
\begin{equation}
\label{TVM}
T_N(\omega)= -N_{\rm eff} \left[(1-\omega)\ln(1-\omega)
  + \omega \ln\omega\right] ~. 
\end{equation}
This is the same form as on the complete graph and the complete bipartite
network, expect for the value of $N_{\rm eff}$.  To compute $N_{\rm eff}$ for
a network with a power-law degree distribution, $n_k \sim k^{-\nu}$, is a
standard exercise in extreme-value statistics~\cite{G87}, and
the final result is
\begin{equation}
\label{VM:HG:SF:CT}
 T_N \propto N_{\rm eff} \sim
\begin{cases}
N  & \nu>3,\\
N^{2(\nu-2)/(\nu-1)} & 2<\nu<3,\\
{\cal O}(1)& \nu<2,
\end{cases}
\end{equation}
with logarithmic corrections in the marginal cases of $\nu=2,3$.  For
$\nu< 3$, consensus arises quickly because $N_{\rm eff}$ is much less than
$N$ when the degree distribution is sufficiently broad.  Here, a few of
high-degree nodes ``control'' many neighboring low-degree nodes, so the
effective number of independent voters is less than $N$.

Applying this same formalism to the IP, the consensus time is
\begin{equation}
\label{TIP}
T_N(\omega_{-1})= -N_{\rm eff} \left[(1\!-\!\omega_{-1})\ln(1\!-\!\omega_{-1}) 
  \!+\! \omega_{-1} \ln\omega_{-1}\right]\,.
\end{equation}
with $N_{\rm eff}=N\mu_1\mu_{-1}$.  For power-law degree networks, $\mu_1$ and
$\mu_{-1}$ can be straightforwardly obtained to give
\begin{equation}
\label{T:IP}
T_N \propto N_{\rm eff} \sim
\begin{cases}
N  & \nu>2,\\
N^{3-\nu} & \nu<2,
\end{cases}
\end{equation}
with again a logarithmic correction for the marginal case $\nu=2$.  Thus the
consensus time in the IP is linear in $N$ for $\nu>2$ and superlinear in $N$
for $\mu<2$.  Consensus arises slowly because of the difficulty in changing
the opinions of agents on the very many low-degree nodes.

\section{BIASED DYNAMICS}
\label{sec:evo}

What happens when the two states are inequivalent?  We may view state
$\mathbf{1}$ as a mutant with fitness $f>1$ that invades a population of
``residents'' in state $\mathbf{0}$, each of which has fitness $f=1$.  What
is the fixation probability, namely, the probability that a single fitter
mutant overspreads the population?  Such fixation underlies many social and
epidemiological phenomena (see e.g.,~\cite{M,K83,AM,E,nowak,PV01,W}).

We implement biased dynamics for the VM as follows:
\begin{itemize}
\itemsep -1ex
\item Pick a voter with probability proportional to its inverse fitness.
\item The voter adopts the state of a random neighbor.
\end{itemize}
Thus a ``weaker'' voter is more likely to be picked and be influenced by
a neighbor.  We may equivalently view the inverse fitness as the death rate
for a given voter.  Similarly, the evolution steps in the biased IP are:
\begin{itemize}
\itemsep -1ex
\item Pick an invader with probability proportional to its fitness.
\item The invader exports its state to a random neighbor.
\end{itemize}
A fitter mutant is thus more likely to spread its progeny.

\begin{figure}[ht]
\centerline{\includegraphics[width=0.4\textwidth]{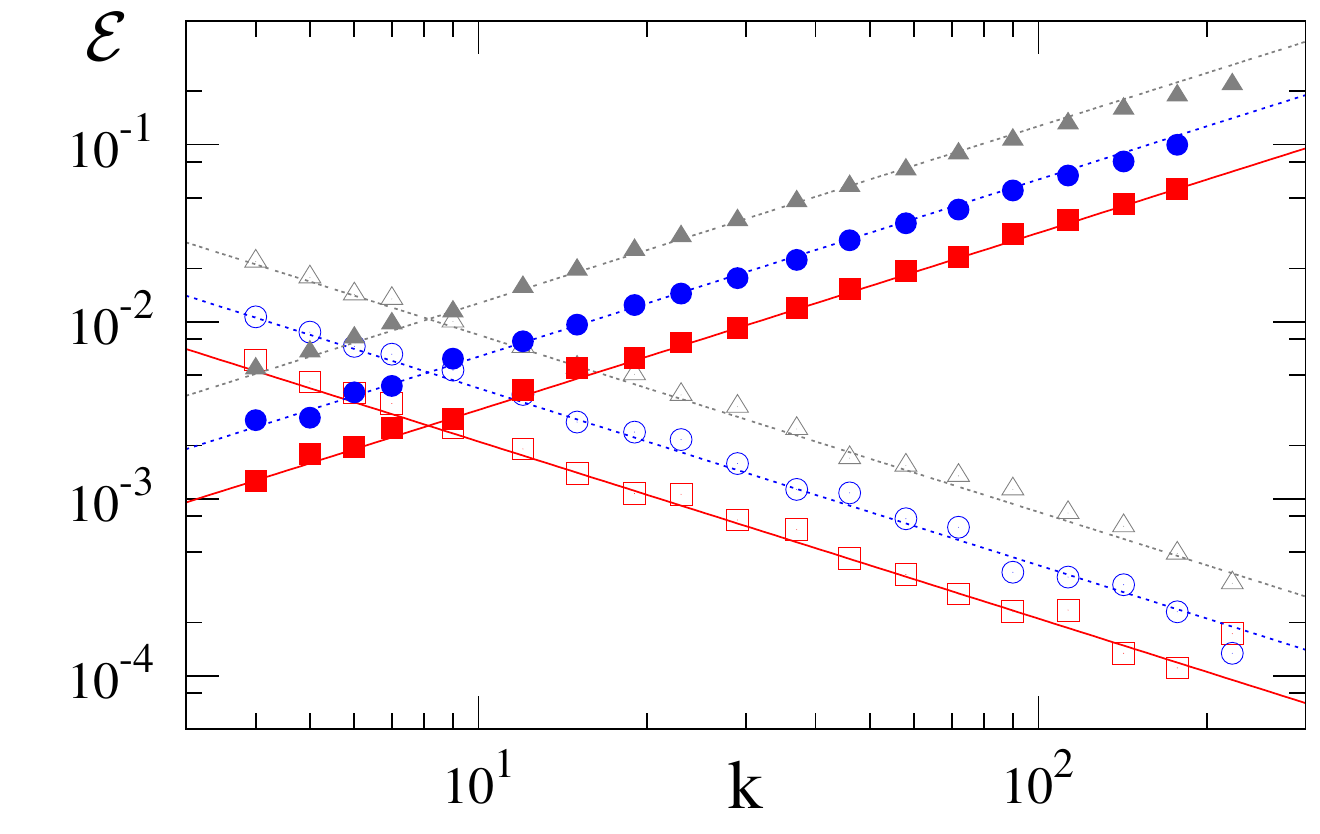}}
\caption{Fixation probability of a single mutant on a node of degree $k$ in
  the configuration model with $n_k \sim k^{-\nu}$ and $\nu = 2.5$, with
  $N = 10^3$ and $\mu_1 = 8$.  Filled symbols correspond to the VM with
  $s = 0.01$, ($\blacksquare$), $s = 0.02$ ($\bullet$) and $s = 0.08$
  ($\blacktriangle$).  Open symbols correspond to IP dynamics with
  $s = 0.004$ ($\square$), $s = 0.008$ ($\circ$) and $s = 0.016$
  ($\triangle$).  The solid lines correspond to the second of
  Eqs.~(\ref{VM:bVM:fix:k:sol}) and (\ref{VM:bIP:fix:k:sol}).}
  \label{fix-k}
\end{figure}

In unbiased dynamics, we saw that high-degree nodes strongly influence the
fixation probability in the VM, while low-degree nodes are more influential
in the IP.  This trend is confirmed by Fig.~\ref{fix-k}, where the fixation
probability is proportional to the degree of the mutant node in the VM and
proportional to the inverse of this degree in the IP.

To understand the fixation probability, let's again consider the simple
example of the complete graph.  The raising and lowering operators in
Eq.~\eqref{rl} now are
\begin{align*}
\begin{split}
  \mathbf{R}(\rho)&\equiv \mathbf{P}[\rho\rightarrow\rho+\delta \rho] = 
\rho(1-\rho) \\
  \mathbf{L}(\rho)&\equiv \mathbf{P}[\rho\rightarrow\rho -\delta \rho] = 
\frac{1}{f}\,\rho(1-\rho),
\end{split}
\end{align*}
We now write the backward Kolmogorov equation for $\mathcal{E}(\rho)$, the
fixation probability to reach consensus when the initial density of agents in
state $\mathbf{1}$ equals $\rho$:
\begin{eqnarray}
\label{Eb}
  \mathcal{E}(\rho)&=   \mathbf{R}(\rho)\mathcal{E}(\rho\!+\!\delta\rho)
 +\mathbf{L}(\rho) \mathcal{E}(\rho\!-\!\delta\rho) \nonumber\\
&+ [1\!-\!\mathbf{R}(\rho)\!-\!\mathbf{L}(\rho)] \mathcal{E}(\rho)\,,
\end{eqnarray}
subject to the boundary conditions $\mathcal{E}(\rho\!=\!0)=0$ and
$\mathcal{E}(\rho\!=\!1)=1$.  In analogy with Eq.~\eqref{T}, this equation
expresses the fixation probability as the appropriately weighted average of
the fixation probabilities after a single update step.  In the following, we
focus on the \emph{weak selection} limit, in which $f=1+s$, with $s\ll 1$.
Expanding \eqref{Eb} to second order in $\delta\rho$ gives
\begin{equation}
\label{VM:bVM:RG:KBG}
\rho(1-\rho) \left[s \frac{\partial\mathcal{E}}{\partial \rho}
+\frac{1}{N}  \frac{\partial^2\mathcal{E}}{\partial \rho^2} \right]=0\,.
\end{equation}
This coincides with the equation for the fixation probability to $\rho=1$ for
biased diffusion on the finite interval $[0,1]$, with solution~\cite{R01,E}
\begin{equation}
\label{bLDsolcont}
\mathcal{E}(\rho;sN)\simeq  \frac{1-e^{-sN\rho}}{1-e^{-sN}}\,.
\end{equation}
Here, we explicitly write the dependence of the fixation probability on
$\rho$ as well as on a second natural variable combination $sN$.

To obtain the fixation probability on a complex network, we extend the two
time-scale dynamics of the unbiased VM to biased dynamics.  Here the
population is again quickly driven to a homogeneous state where
$\rho_k\to \omega$ for all $k$ on a time scale of the order of 1.  Once this
homogeneous state is reached, the new feature is that consensus is driven by
the bias, rather than by diffusive fluctuations.  Thus we are led to study
the evolution of $\langle\omega\rangle$, which, for $s>0$, evolves as
$ \langle\dot\omega\rangle= s\langle\omega\rangle (1-\langle\omega\rangle)$.
This gives $\langle\omega\rangle \to 1$ on a time scale of the order of
$s^{-1}\gg 1$.

We now determine the fixation probability by applying the same computational
approach as that for the unbiased VM: replace $\rho_k$ by $\omega$ in all
transition probabilities and the derivative
$\frac{\partial}{\partial \rho_k}$ by
$\frac{kn_k}{\mu_1}\frac{\partial}{\partial \omega}$.  With these
replacements, the backward Kolmogorov equation for the fixation probability
has the same form as Eq.~\eqref{VM:bVM:RG:KBG}, but with $N$ replaced by
$N_{\rm eff}=N\mu_1^2/\mu_2$ and $\rho$ by $\omega$.  The fixation
probability for biased dynamics on a complex network is then given by
Eq.~\eqref{bLDsolcont} with these replacements.

For a single mutant initially at a node of degree $k$, $\omega = k/N\mu_1$.
Substituting this into~\eqref{bLDsolcont}, we fine generally that the
fixation probability is proportional to $k$ for all $s\ll 1$ and has the
limiting behaviors (Fig.~\ref{fix-k}):
\begin{equation}
\label{VM:bVM:fix:k:sol} 
\mathcal{E}  \simeq
\begin{cases}
    {\displaystyle k/N\mu_1} & \quad s \ll 1/N_{\mathrm{eff}} ;\\
       {\displaystyle  k\, (s \mu_1/\mu_2)} & \quad  1/N_{\mathrm{eff}} \ll s \ll 1.
\end{cases}
\end{equation} 

In the complementary biased IP, the fixation probability for a mutant
initially on a node of degree $k$ is  inversely
  proportional to the node degree:
\begin{equation}
  \label{VM:bIP:fix:k:sol} 
  \mathcal{E} \simeq
\begin{cases}
{\displaystyle k^{-1} /N\mu_{-1}} &\quad s\ll 1/N ;\\
 {\displaystyle k^{-1}(s/\mu_{-1})}  &\quad 1/N\ll s\ll 1\,.
\end{cases} 
\end{equation}

\vspace{-0.25cm}
\section{SUMMARY}
\vspace{-0.2cm}

The venerable voter model played a central role in probability theory and
statistical physics because it is one of the few exactly soluble
many-particle interacting systems in all spatial dimensions and because of
the diversity of its applications.  Putting the voter model on a complex
network---in which there is broad distribution of node degrees---changes its
dynamics in crucial ways.

A new dynamical conservation law---the \emph{degree-weighted}
magnetization---gives the fixation probability for the voter model and the
invasion process on finite networks.  Another new feature is a two time-scale
approach to consensus---first an initial quick approach to a homogeneous
state in which the density of $\mathbf{1}$s is the same for nodes of any
degree, after which diffusive fluctuations drive the consensus.  Consensus is
achieved quickly in the voter model when the degree distribution is
sufficiently broad, as high-degree nodes effectively ``control'' many
neighboring low-degree nodes.  When one state is more fit, there is again a
two time-scale approach to consensus, but with fitness selection driving
ultimate consensus.  As a message for evolutionary dynamics, for a mutant to
infiltrate a network most effectively, it is advantageous for it to be on a
high-degree node in the voter model and on a low-degree node in the invasion
process.


\begin{thebibliography}{99}

\bibitem{SAR} V. Sood and S. Redner, Phys.\ Rev.\ Lett.\ {\bf 94}, 178701
  (2005); T. Antal, S. Redner, and V. Sood, Phys.\ Rev.\ Lett.\ {\bf
    96}, 188104 (2006); V. Sood, T. Antal, and S. Redner, Phys.\ Rev.\ E {\bf
    77}, 041121 (2008).

\bibitem{L} T. M. Liggett, \emph{Interacting Particle Systems},
  (Springer-Verlag, Berlin, 2005).

\bibitem{K92} P. L. Krapivsky, Phys.\ Rev.\ A {\bf 45}, 1067 (1992).

\bibitem{VK} N. G. van Kampen, \emph{Stochastic Processes in Physics and
    Chemistry}, $2^{\rm nd}$ ed.\ (North-Holland, Amsterdam, 1997).

\bibitem{R01} S. Redner, \emph{A Guide to First-Passage Processes}, (Cambridge
  University Press, New York, 2001).

\bibitem{N} M. E. J. Newman, \emph{Networks: An Introduction} (Oxford
  University Press, 2010).

\bibitem{G87} J. Galambos, \emph{The Asymptotic Theory of Extreme Order
Statistics}, (R. E. Krieger Publishing Co., Malabar, Florida, 1987).

\bibitem{M} P. A. P Moran, \emph{The Statistical Processes of Evolutionary
    Theory} (Clarendon Press, Oxford, 1962).

\bibitem{K83} M. Kimura, \emph{The Neutral Theory of Molecular Evolution},
  (Cambridge University Press, Cambridge, 1983).

\bibitem{AM} R. M. Anderson and R. M. May, \emph{Infectious Diseases in
    Humans}, (Oxford University Press, Oxford, 1992).

\bibitem{E} W. Ewens, \emph{Mathematical Population Genetics I. Theoretical
  Introduction}, (Springer-Verlag, Berlin, 2004).

\bibitem{nowak} M. A. Nowak, \emph{Evolutionary Dynamics}, (Harvard Univ.\
  Press, Cambridge MA 2006).

\bibitem{PV01} R. Pastor-Satorras and A. Vespignani, Phys.\ Rev.\ Lett.\ {\bf
    86}, 3200 (2001).

\bibitem{W} D. J. Watts, Proc.\ Natl.\ Acad.\ Sci.\ USA {\bf 99}, 5766 (2002).


\end{thebibliography}
\end{document}